\documentclass[conference]{IEEEtran}
\IEEEoverridecommandlockouts

\usepackage{cite}
\usepackage{amsmath,amssymb,amsfonts}
\usepackage{algorithmicx}
\usepackage{graphicx}
\usepackage{textcomp}
\usepackage{xcolor}

\usepackage{amsmath}
\usepackage{graphicx} 
\usepackage{multirow}  
\usepackage{booktabs}  
\usepackage{array}    
\usepackage{caption} 
\usepackage{colortbl} 
\usepackage{xcolor}   
\usepackage{tabularx}
\usepackage{tikz}
\usepackage{amsmath}
\usepackage{array}
\usepackage{booktabs}
\usepackage{multirow}
\usepackage{tabularx}
\usepackage{adjustbox}
\usepackage{makecell}
\usepackage{caption}
\usepackage{xspace}
\usepackage{pifont}
\usepackage{amssymb}
\usepackage{tablefootnote}
\usepackage{filecontents}
\usepackage{threeparttable}
\usepackage{subfigure}
\usepackage{algorithm}
\usepackage{algpseudocode}
\usepackage{amsmath} 
\usepackage{xurl}
\usepackage{hyperref}

\newcommand{\etal}{{\textit{et al.}}}

\def\BibTeX{{\rm B\kern-.05em{\sc i\kern-.025em b}\kern-.08em
    T\kern-.1667em\lower.7ex\hbox{E}\kern-.125emX}}
\begin{document}

\title{Boosting the Transferability of Audio Adversarial Examples with Acoustic Representation Optimization}

\author{Weifei Jin, Junjie Su, Hejia Wang, Yulin Ye, Jie Hao$^*$\thanks{$^*$Corresponding author.} \\  
\em \small National Engineering Research Center of Disaster Backup and Recovery\\  
School of Cyberspace Security, Beijing University of Posts and Telecommunications \\
\rm Beijing, China \\
\rm \small \texttt{\{weifeijin, sujunjie, hk1182626120stan, yyl5632, haojie\}@bupt.edu.cn}
}

\maketitle

\begin{abstract}
With the widespread application of automatic speech recognition (ASR) systems, their vulnerability to adversarial attacks has been extensively studied. However, most existing adversarial examples are generated on specific individual models, resulting in a lack of transferability. In real-world scenarios, attackers often cannot access detailed information about the target model, making query-based attacks unfeasible. 
To address this challenge, we propose a technique called Acoustic Representation Optimization that aligns adversarial perturbations with low-level acoustic characteristics derived from speech representation models. Rather than relying on model-specific, higher-layer abstractions, our approach leverages fundamental acoustic representations that remain consistent across diverse ASR architectures. By enforcing an acoustic representation loss to guide perturbations toward these robust, lower-level representations, we enhance the cross-model transferability of adversarial examples without degrading audio quality.
Our method is plug-and-play and can be integrated with any existing attack methods. We evaluate our approach on three modern ASR models, and the experimental results demonstrate that our method significantly improves the transferability of adversarial examples generated by previous methods while preserving the audio quality.

\end{abstract}

\begin{IEEEkeywords}
audio adversarial example, acoustic representation, transfer attack
\end{IEEEkeywords}

\section{Introduction}
With the increasing deployment of Automatic Speech Recognition (ASR) systems in a variety of settings, ranging from voice-driven personal assistants to healthcare diagnostics and transcription services~\cite{latif2020speech,dhanjal2024comprehensive,liao2023improving}, their security vulnerabilities have attracted mounting scrutiny. Among these, the threat of adversarial perturbations has emerged as a particularly pressing concern~\cite{ge2023advddos,zhang2024mitigating,jin2024towards}. Such perturbations, which are often imperceptible to human listeners, can be injected into audio inputs to mislead ASR models into generating incorrect or even malicious transcriptions~\cite{carlini2018audio}.

However, a significant limitation of most existing attack methods, whether gradient-based (white-box) or query-based (black-box), is their strong dependency on a single target model~\cite{qu2022synthesising,yu2023smack}. As a result, the resulting adversarial examples (AEs) tend to exhibit poor transferability when applied to other models. This shortcoming becomes critical in realistic black-box scenarios, where adversaries typically lack any knowledge of the target model’s architecture or parameters. The inability to produce robust, transferable AEs under such conditions severely restricts the practicality of existing attack strategies. Consequently, enhancing transferability to effectively deceive unseen ASR systems has become a central challenge in the field of adversarial audio attacks.

\begin{figure}[t]
    \centering\includegraphics[width=0.99\linewidth]{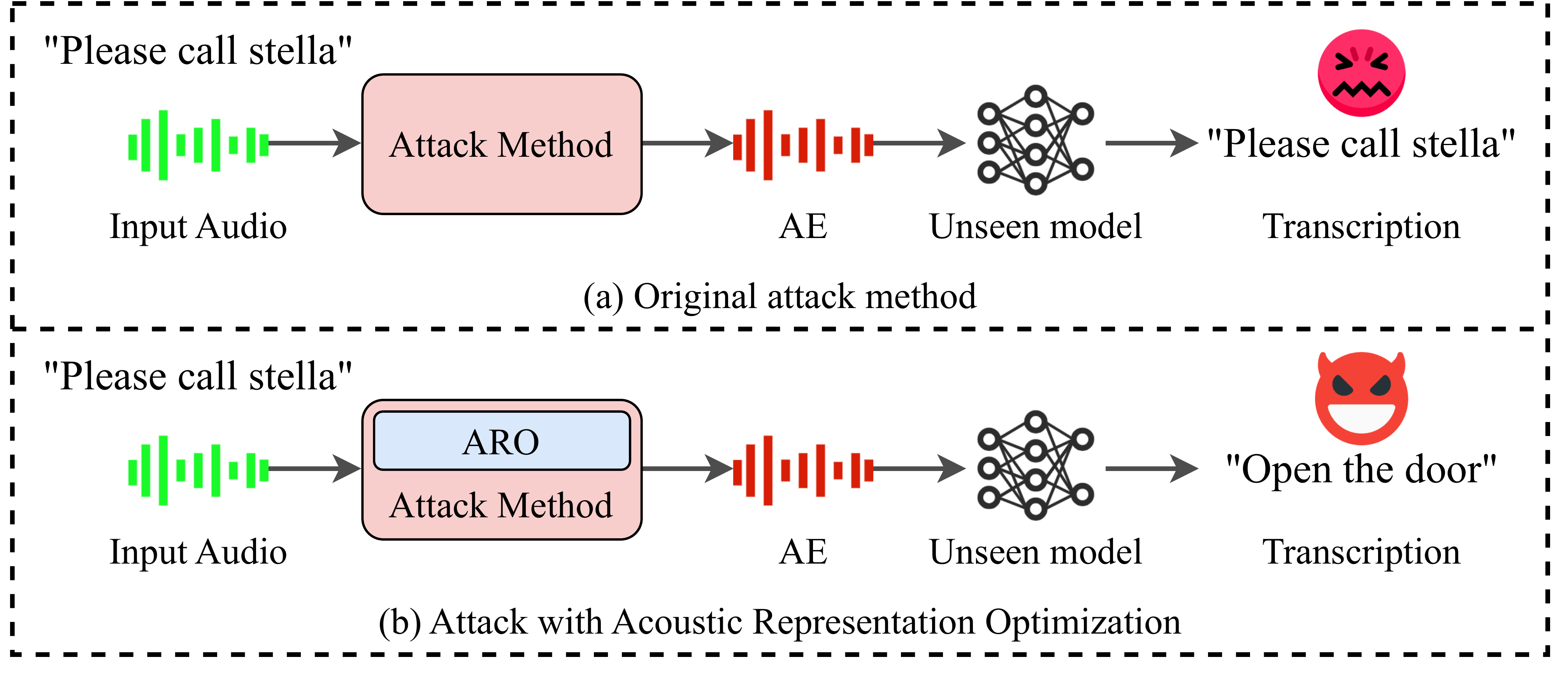}
    \caption{Illustration of the transferability of the attack method with/without acoustic representation optimization.}
    \label{fig:overview}
    \vspace{-4mm}
\end{figure}

To address this challenge, we propose an \textbf{A}coustic \textbf{R}epresentation \textbf{O}ptimization (ARO) technique that leverages the low-level acoustic representations extracted from speech representation models (SRMs). Instead of focusing on model-specific, higher-layer representations, we guide adversarial examples to align with fundamental acoustic representations, such as spectral patterns, drawn from the lower layers of an SRM. In doing so, we reduce the risk of overfitting to a particular model’s decision space, thereby enhancing the transferability of the generated adversarial examples across unseen ASR systems.
Concretely, we employ a carefully designed loss function based on the cosine similarity between the acoustic representations of the adversarial examples and those of a chosen target audio. By encouraging perturbations to preserve or emphasize these basic acoustic signatures, our method ensures that the resulting adversarial examples remain effective even when confronted with different ASR architectures. Moreover, our acoustic representation optimization integrates seamlessly into existing attack pipelines, adding minimal overhead and preserving the audio quality. As shown in Figure~\ref{fig:overview}, after incorporating our method, existing attack methods achieve better attack performance on unseen models. Our empirical analyses further validate that low-level acoustic representations exhibit greater generalizability and contribute more effectively to the transferability of adversarial audio examples than their high-level counterparts. As a result, our approach not only elevates the cross-model attack success rate but also improves the practicality and robustness of adversarial attacks in real-world black-box scenarios.

Experiments on two commercial ASR APIs and one leading open-source ASR model demonstrate that adversarial examples crafted using our acoustic representation optimization technique achieve significant improvements in cross-model transferability, while preserving audio quality. In comparison to previous adversarial attacks, many of which falter when confronted with heterogeneous ASR architectures, our approach maintains robust attack performance across a diverse range of models. This heightened adaptability underscores the method’s practical potential for real-world black-box scenarios.

In summary, our contributions are fourfold:
\begin{itemize}
    \item We introduce an acoustic representation optimization strategy that guides adversarial perturbations to align with low-level acoustic attributes, thereby improving the transferability of adversarial examples.
    \item We empirically demonstrate that the acoustic features derived from the lower-layer representations of SRMs exhibit greater generalizability compared to those from higher-layer representations, thereby facilitating enhanced transferability of adversarial audio examples.
    \item Our method seamlessly integrates into existing adversarial attack methods without modifying their original optimization processes, ensuring ease of adoption and practicality.
    \item We conduct extensive experiments on multiple ASR systems, confirming that our approach significantly enhances attack success rates while maintaining audio quality, thus providing strong empirical evidence of its effectiveness in realistic conditions.
\end{itemize}

\begin{figure*}[t]
    \centering\includegraphics[width=0.68\linewidth]{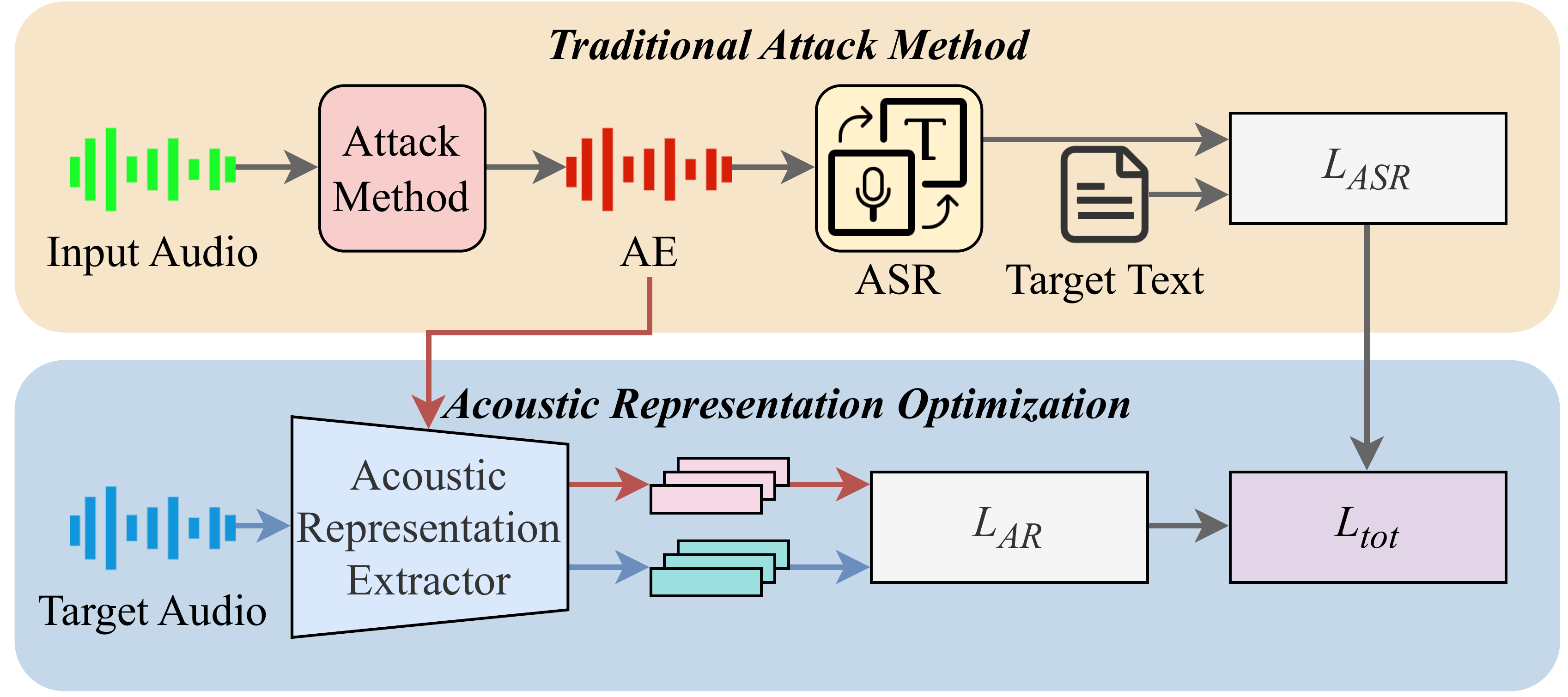}
    \caption{Overview of acoustic feature optimization integrated with traditional attack methods.}
    \label{fig:method}
\end{figure*}

\section{Related Work}

\noindent\textbf{Adversarial Attacks on ASR Models.}
Deep neural networks (DNNs) have been widely used in ASR systems, which also exposes them to the threat of adversarial attacks~\cite{szegedy2013intriguing}. Carlini \etal~\cite{carlini2018audio} first proposed an optimization method, C\&W, in the audio field under the white-box setting, achieving attacks on Mozilla DeepSpeech2~\cite{amodei2016deep}. Yakura \etal~\cite{yakura2018robust} introduced the Y\&S method to address the gap where C\&W is not robust against physical world transformations when propagating over the air. To further improve the efficiency, FAAG~\cite{miao2022faag} was proposed to accelerate previous methods by reducing computational cost. Finally, SSA~\cite{qu2022synthesising} was proposed to handle the situation when the original audio is unavailable, by directly synthesizing adversarial examples in a text-to-speech (TTS) model.
However, these attacks are limited to white-box scenarios and weakened in realistic black-box scenarios. Chen \etal~\cite{chen2020devil} proposed Devil's Whisper, which generates a surrogate model locally by querying the black-box target model, achieving attacks on commercial ASR systems including Google Assistant, Google Home, Microsoft Cortana, and Amazon Echo. However, such attacks still require a large number of queries, which is cost-expensive.

\noindent\textbf{Transferable Adversarial Attacks.}
To break the limitations of query-based black-box attacks, transfer attacks exploit the transferability of adversarial examples among different models. Transferable adversarial examples are trained on a carefully chosen surrogate model. However, the overfitting problem of optimizing adversarial examples on the surrogate model resulting in loss of attack effectiveness on unseen models. Qi \etal~\cite{qi2023transaudio} proposed Transaudio, a novel contextualized attack method that addresses the overfitting problem and enables arbitrary word-level attacks through the incorporation of deletion, insertion, and substitution adversarial behaviors. In order to further enhance transferability, the ensemble method~\cite{hang2020ensemble} was introduced later to attack ASR systems~\cite{zhang2021generating, fang2024zero}. Since multiple surrogate models are used for training, the generated adversarial perturbations can learn more common features among different models. However, these methods also suffer from the problem of high computational cost. Unlike prior transfer-based approaches, our method is designed to be plug-and-play, requiring no architectural changes, which significantly improves its applicability in diverse attack settings.

\section{Method}
\subsection{Design Intuition}
One of the main challenges in generating transferable adversarial examples for ASR is that these perturbations often overfit to the decision space of a single model. Traditional adversarial attacks typically rely on the final layer’s output (often optimized via connectionist temporal classification, CTC), making the adversarial examples highly model-specific. As a result, perturbations crafted on one model frequently fail to deceive another, limiting their transferability.

A natural approach to enhance cross-model transferability is to focus on the more fundamental, model-agnostic aspects of speech, rather than model-dependent representations. Initially, one might hypothesize that intermediate or higher-layer representations of SRMs would offer an advantage, as they seem to capture richer linguistic and semantic structures. However, our empirical findings suggest the opposite: extracting features from \emph{lower} layers of SRMs consistently yields better transfer performance. This observation indicates that the basic acoustic patterns present at lower-level representations are more stable and consistent across a range of ASR architectures.

In contrast, higher-layer representations, despite being more semantically rich, tend to be more intertwined with the specific training strategies, architectures, and objectives of particular models. Hence, adversarial perturbations tuned to these higher-level features risk overfitting and failing to generalize. In contrast, aligning the adversarial perturbations with low-level acoustic representations (such as spectral envelopes and basic time-frequency patterns) anchors the perturbations to aspects of the speech signal that are less sensitive to the internal variations of different models.

In summary, our design intuition is based on the insight that \emph{lower-level} SRM representations capture fundamental acoustic attributes that are less influenced by model-specific biases. By guiding adversarial examples to align with these stable acoustic representations, we improve the inherent transferability of the generated perturbations.

\subsection{Acoustic Representation Optimization}
\noindent\textbf{Overview.}
Our goal is to enhance the transferability of adversarial examples generated by existing attack methods while maintaining audio quality. Figure~\ref{fig:method} illustrates the overall framework of our method. Previous white-box attack methods typically conduct targeted attacks on a surrogate model locally. Without modifying the workflow of traditional attack methods, we simply incorporate our proposed acoustic representation optimization process during the local optimization. After generating the adversarial examples, they are transferred to the target black-box model for testing. The pseudocode of the entire algorithm is presented in Algorithm~\ref{alg:SFO}. 
Specifically, the optimization process follows these steps:

1) Select one layer from a model among several SRMs and layers as the acoustic feature extractor $\mathcal{E}(\cdot)$.

2) For the current audio sample \(x\) and the given target text \(t\), choose the audio \(x_t\) corresponding to the target text as the target audio, and extract the acoustic feature \(s_{t} = \mathcal{E}(x_t)\).

3) In each iteration, extract the acoustic representation of the current adversarial example \(x_{adv}\) using the acoustic representation extractor, \(s_{adv} = \mathcal{E}(x_{adv})\), and compute the cosine similarity \(\text{Cos\_Sim}(s_{adv}, s_t)\) between the acoustic representations of the adversarial example and the target audio. \(\mathcal{L}_{AR} = 1 - \text{Cos\_Sim}(s_{adv}, s_t)\) is calculated as the acoustic representation loss.

4) Combine the acoustic representation loss with the original loss of the existing method and optimize the adversarial example using gradient descent.

\begin{algorithm}[!t]
\caption{Acoustic Representation Optimization (ARO)}
\label{alg:SFO}
\renewcommand{\algorithmicrequire}{\textbf{Input:}}
\renewcommand{\algorithmicensure}{\textbf{Output:}}
\begin{algorithmic}[1]
\Require acoustic representation extractor $\mathcal{E}$, local ASR model $f$, target text $t$, corresponding target audio $x_t$, scaling factor $\beta$, maximum iteration number $N$, learning rate $\alpha$, perturbation threshold $\tau$, ASR loss $\mathcal{L}_{ASR}$, acoustic representation loss $\mathcal{L}_{AR}$.
\Ensure Optimized adversarial example $x^*$.
\State Randomly initialize $\delta$
\State $s_t\gets \mathcal{E}(x_t)$ 
\For{$i \leftarrow 1$ \textbf{to} $N$}
    \State Get current AE $x_{adv}$ based on $\delta$
    \State Calculate original loss $\mathcal{L}_{adv}$
    \State $s_{adv} \gets \mathcal{E}(x_{adv})$
    \State Calculate acoustic representation loss $\mathcal{L}_{AR}$
    \State $\mathcal{L}_{tot}\gets\mathcal{L}_{adv}+\beta\cdot\mathcal{L}_{AR}$
    \State $\delta \gets \delta -\alpha \cdot \text{sign}(\nabla_{\delta}\mathcal{L}_{tot}(x,\delta,t,x_t,f))$
\EndFor
\State $\delta^*\gets \delta$, $x^*\gets x+\delta^*$
\State \Return $x^*$
\end{algorithmic}
\end{algorithm}

\noindent\textbf{Optimization Process.}
Given an audio sample $x$ and a target text $t$, the traditional white-box adversarial attack method typically has the following optimization objective for a local surrogate ASR model $f$:
\begin{equation}
\begin{aligned}
\delta^* &= \mathop{\text{argmin}}_{\delta} \mathcal{L}_{adv}(x, \delta, t, f) \\
    &= \mathop{\text{argmin}}_{\delta} \mathcal{L}_{ASR}(f(x + \delta), t) + \lambda \mathcal{L}_{reg}(\delta) \\
\text{s.t.}& \quad \mathcal{Q}(\delta) \leq \tau,
\end{aligned}
\end{equation}
where $\delta$ is the adversarial perturbation to be optimized, $\mathcal{L}_{reg}$ is the regularization loss controlled by the weight coefficient $\lambda$, typically $\|\delta\|_2^2$. $\mathcal{Q}(\cdot)$ is the quality function that constrains the quality loss within the threshold $\tau$. For example, in C\&W~\cite{carlini2018audio}, the distortion of the audio signal is used to evaluate the quality loss of speech:
\begin{equation}
\begin{aligned}
\mathcal{Q}(\delta) =& \, \text{dB}(\delta) - \text{dB}(x) \\
\text{dB}(x) =& \max_i \, 20 \cdot \log_{10}(x_i).
\end{aligned}
\end{equation}

For acoustic representation optimization, given a acoustic representation extractor $\mathcal{E}$ and a target audio $x_t$ corresponding to the target text $t$, we aim to optimize the following loss function:
\begin{equation}
    \mathcal{L}_{AR}(x, \delta, x_t) = 1 - \text{Cos\_Sim}(\mathcal{E}(x + \delta), \mathcal{E}(x_t)),
\end{equation}
where $\text{Cos\_Sim}(\cdot,\cdot)$ denotes the cosine similarity, with a range of $\left[-1, 1 \right]$. Thus, the final total loss function is:
\begin{equation}
    \mathcal{L}_{tot}(x, \delta, t, x_t, f) = \mathcal{L}_{adv}(x, \delta, t, f) + \beta \cdot \mathcal{L}_{AR}(x, \delta, x_t),
\end{equation}
where $\beta$ is a scaling factor that controls the balance between the two loss functions. Similar to the original attack method, we use I-FGSM~\cite{kurakin2018adversarial} for optimization:
\begin{equation}
    \delta' = \delta - \alpha \cdot \text{sign}(\nabla_{\delta} \mathcal{L}_{tot}(x, \delta, t, x_t, f)),
\end{equation}
where $\alpha$ is the learning rate.

\section{Experiments}
\begin{table*}[t]
\centering
\caption{Comparison of three existing methods with/without ARO on four ASRs (DeepSpeech, Alibaba, iFlytek, and Whisper).}
\resizebox{0.99\linewidth}{!}{
\begin{tabular}{c|c c c|c c c|c c c|c c c|c|c}
\Xhline{1px}
\multirow{2}{*}{\textbf{Method}} 
& \multicolumn{3}{c|}{\textbf{DeepSpeech}} 
& \multicolumn{3}{c|}{\textbf{Alibaba}} 
& \multicolumn{3}{c|}{\textbf{iFlytek}} 
& \multicolumn{3}{c|}{\textbf{Whisper}} 
& \multirow{2}{*}{\textbf{SNR}} 
& \multirow{2}{*}{\textbf{MOS}} \\ \cline{2-13}
& \textbf{SRoA} & \textbf{CER} & \textbf{WER} 
& \textbf{SRoA} & \textbf{CER} & \textbf{WER} 
& \textbf{SRoA} & \textbf{CER} & \textbf{WER} 
& \textbf{SRoA} & \textbf{CER} & \textbf{WER} 
& & \\ \Xhline{1px}
C\&W~\cite{carlini2018audio} & 100.00 & 84.79 & 113.11 & 12.50 & 17.38 & 29.20 & 26.00 & 28.55 & 44.64 & 29.50 & 44.36 & 51.41 & 8.84 & 1.71 \\ \hline
\rowcolor{gray!20} C\&W+ARO & 100.00 & 83.98 & 110.02 & 38.50 & 42.15 & 58.06 & 62.00 & 56.26 & 75.61 & 60.00 & 55.63 & 77.05 & 8.83 & 1.72 \\ \hline
Y\&S~\cite{yakura2018robust} & 95.00 & 92.70 & 143.06 & 22.00 & 27.91 & 42.09 & 36.50 & 41.56 & 65.25 & 38.50 & 43.57 & 59.67 & 7.62 & 1.87 \\ \hline
\rowcolor{gray!20} Y\&S+ARO & 97.00 & 98.54 & 162.63 & 40.50 & 44.19 & 61.48 & 64.50 & 60.86 & 80.64 & 60.00 & 55.69 & 78.04 & 8.06 & 1.87 \\ \hline
FAAG~\cite{miao2022faag} & 100.00 & 85.30 & 112.43 & 5.00 & 12.40 & 21.85 & 20.50 & 26.66 & 42.42 & 10.50 & 20.93 & 31.37 & 15.00 & 2.71 \\ \hline
\rowcolor{gray!20} FAAG+ARO & 100.00 & 84.31 & 111.03 & 27.50 & 32.08 & 45.63 & 41.00 & 42.90 & 60.67 & 42.00 & 46.78 & 63.58 & 13.07 & 1.92 \\ \Xhline{1px}
\end{tabular}
}
\vspace{-5mm}
\label{tab:asr_comparison}
\end{table*}

\subsection{Experiment Setup}
\noindent\textbf{Datasets and Settings.} We randomly select 200 audio clips from the commonly used LibriSpeech dataset~\cite{panayotov2015librispeech} to construct adversarial examples. Additionally, we randomly choose 50 text-audio pairs from the VCTK Corpus~\cite{yamagishi2019cstr}. For each test sample, we randomly select one pair from the 50 text-audio pairs as the target text and target audio. Ultimately, we obtain 200 audio-target text-target audio triples as the test dataset.We use the Adam~\cite{kingma2014adam} optimizer for optimization, with the learning rate $\alpha$ set to 0.001 and $\beta$ set to 180.

\noindent\textbf{Target Models and SRMs.} Following previous work~\cite{carlini2018audio, qu2022synthesising}, we use DeepSpeech2~\cite{amodei2016deep} as the target model in a white-box setting to optimize the generation of adversarial examples locally. During the testing phase, we perform transfer attacks on two commercial ASRs, Alibaba~\cite{alibaba} and iFlytek~\cite{iflytek}, as well as the state-of-the-art open-source ASR, Whisper-large-v3~\cite{radford2023robust}. In addition, we also evaluate three speech representation models to extract acoustic representations: Wav2vec2~\cite{baevski2020wav2vec}, HuBERT~\cite{hsu2021hubert}, and WavLM~\cite{chen2022wavlm}.

\noindent\textbf{Evaluation Methods.} To assess the plug-and-play nature of our method, we combine it with three existing attack methods, C\&W~\cite{carlini2018audio}, Y\&S~\cite{yakura2018robust} and FAAG~\cite{miao2022faag}. C\&W is a classic attack method that directly adds perturbations in the original acoustic space and optimizes using CTC loss. 
Y\&S introduces an adversarial attack that incorporates a robust objective function to enhance the model’s performance under physical attack. FAAG, on the other hand, is a feature-aware adversarial attack method that uses feature map manipulation to generate more effective perturbations. 
Evaluating our approach on these three methods thoroughly validates the plug-and-play nature of our method.

\noindent\textbf{Metrics.} We use the following five metrics to evaluate the attack performance and audio quality:
\begin{itemize}
    \item SRoA (Success Rate of Attack). We consider an attack successful when more than 50\% of the characters in the transcription differ from the original text in untargeted attacks on black-box models.
    \item CER (Character Error Rate) and WER (Word Error Rate). They measure the percentage of character or word errors in transcriptions compared to the original text. 
    \item SNR (Signal-to-Noise Ratio).  SNR denotes the ratio of the strength of a signal to the level of background noise, and is often used to measure signal clarity. A higher SNR indicates that the signal is stronger relative to the noise, resulting in better quality.
    \item MOS (Mean Opinion Score). The sound quality MOS in this paper is derived from NISQA~\cite{mittag2021nisqa}, a state-of-the-art DNN-based speech quality evaluation system that quantifies overall audio quality and naturalness with scores ranging from 1 to 5.
\end{itemize}

\subsection{Results Analyses}
\noindent\textbf{Evaluation of Attack Performance.}
Results in Table~\ref{tab:asr_comparison} show that after incorporating acoustic representation loss, three methods demonstrate significant improvements in attack performance, with FAAG's attack success rate increasing an average of 283\% of the original attack success rate. The improvements on Alibaba API are particularly notable for all three methods, with success rates exceeding 208\% , 84\% and 450\% of original success rates, respectively. 
Moreover, CER and WER also show substantial improvements. For example, the three methods improved the average CER by 68.14\% and 58.91\% on the iFlytek API and Whisper, respectively. These results indicate that, through our acoustic representation optimization, adversarial examples avoid overfitting to the decision space of the local substitute model and instead align with the stable acoustic representation of the target audio. This helps adversarial examples move further away from the decision boundary in the decision spaces of different models, ultimately improving transferability.

\noindent\textbf{Evaluation of Audio Quality.}
When applying acoustic representation loss, our approach does not significantly degrade the audio quality, as this would reduce the practicality of our method. To evaluate the impact of our approach on audio quality, we conduct comprehensive tests. We use the traditional SNR to assess noise level and adopt the state-of-the-art DNN-based audio quality evaluation system, NISQA, to obtain MOS scores. NISQA evaluates audio quality and naturalness based on four speech quality dimensions: noisiness, coloration, discontinuity, and loudness. 
The evaluation results shown in Table~\ref{tab:asr_comparison} demonstrate that our method does not significantly reduce audio quality. Compared to the original methods, after incorporating acoustic representation optimization, the SNR of C\&W and FAAG only decreased by 0.01 and 1.93, while the SNR of Y\&S actually improved by 0.44. Similarly, the MOS for C\&W and Y\&S changed very little, while the MOS for FAAG decreased slightly by 0.79. This indicates that our acoustic representation loss does not introduce significantly larger perturbations compared to the original methods. Overall, these results show that our method has only a minimal impact on audio quality, which ensures its usability.

\begin{figure}[t]
    \centering
    \includegraphics[width=0.8\linewidth]{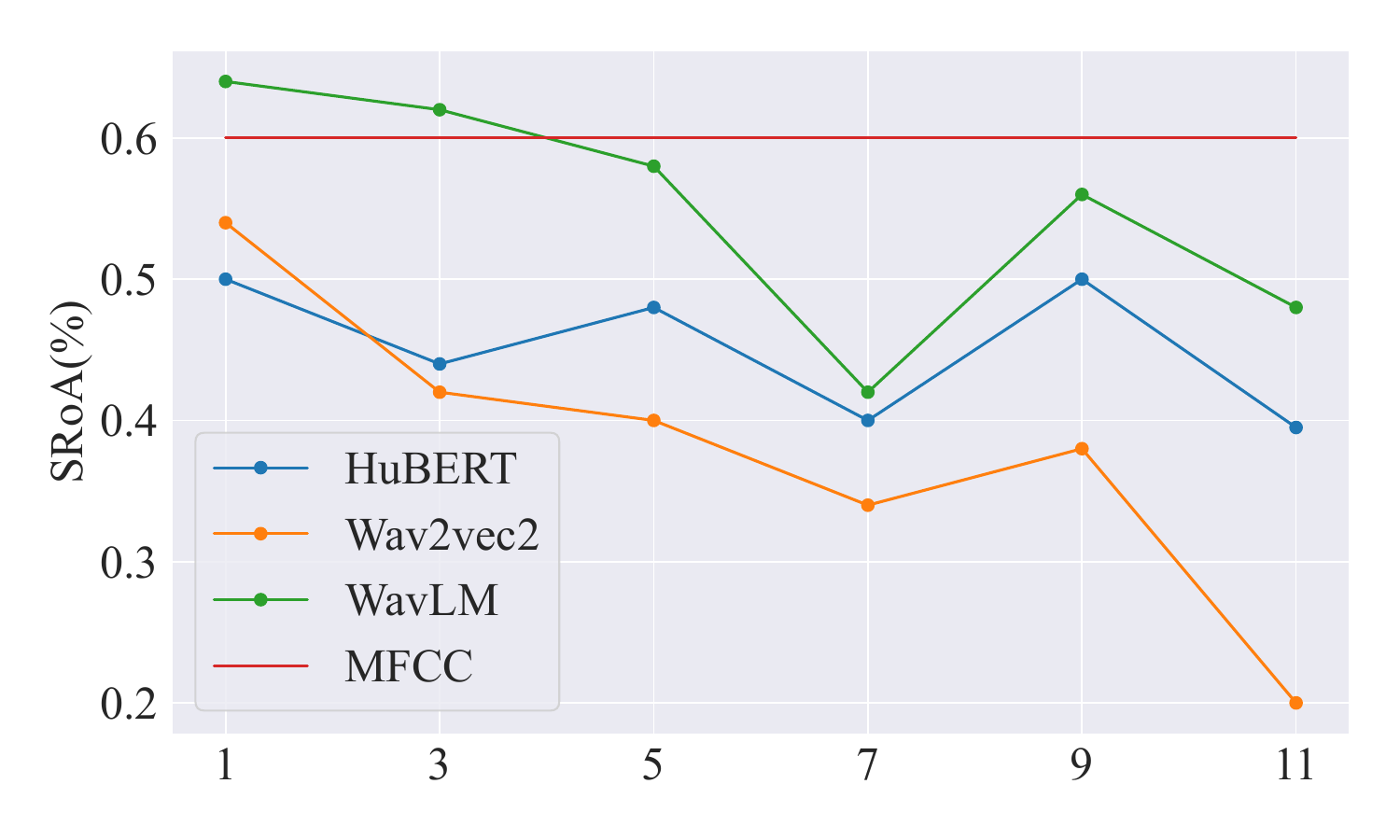}
    \caption{The variation of SRoA with the depth of representation layers in different models.}
    \label{fig:layers}
    \vspace{-7mm}
\end{figure}

\noindent\textbf{Evaluation on Different SRMs.}
Wav2vec2~\cite{baevski2020wav2vec}, HuBERT~\cite{hsu2021hubert}, and WavLM~\cite{chen2022wavlm}.
The feature representations extracted by speech representation models often contain various types of information, such as acoustic features, semantic features, speaker features, and more, while our primary focus is on acoustic representations. To obtain the best acoustic representations, we evaluate several state-of-the-art SRMs known for their performance in speech feature extraction, and we also provided MFCC features as a reference.
As shown in Figure~\ref{fig:layers}, using WavLM~\cite{chen2022wavlm} to extract acoustic representations results in the best improvements in transferability for adversarial attacks, with improvements of 117\% (relative values) for the C\%W methods on Whisper compared to the original methods. This highlights WavLM's superior ability to extract acoustic representations.
The performance of Wav2Vec2~\cite{baevski2020wav2vec} and HuBERT~\cite{hsu2021hubert} is not as good. We suspect this is because their design aims to extract semantic features, which interferes with the extraction of acoustic features. Specifically, Wav2Vec2 gradually transforms speech signals into more abstract representations through its continuous feature encoder, which makes it perform better in tasks such as speech recognition and semantic understanding. HuBERT, in particular, focuses on uncovering latent linguistic structures in speech through clustering. On the other hand, WavLM not only extracts general acoustic and semantic features using self-supervised learning but also incorporates objectives from multiple tasks (such as speech recognition and speaker identification). This enables it to capture both semantic information and other low-level features, allowing its lower-layer representations to perform better in extracting low-level acoustic features.
In summary, we select WavLM as the acoustic representation extractor.

\noindent\textbf{Evaluation of Different Layers.}
In general, the higher-level features of SRM mainly represent high-level semantic information, while the shallow layers extract local acoustic features. To select the best layer for acoustic representation extraction, we also evaluate different layers of SRM using the C\&W method, with Whisper for recognition, and present the results in Figure~\ref{fig:layers}.
An important finding is that as the layer depth increases, the improvement in feature representations for transfer learning gradually weakens. This result aligns with our design intuition, which suggests that the lower-level features of the model are more beneficial for adversarial transferability. This indicates that in SRMs, lower-level acoustic features tend to adapt better for transfer across different models, while higher-level features may become overly model-dependent, leading to poor performance when transferred to a new model.
We initially assume that the goal of ASR models is to transcribe speech signals into text, so the higher-level features of the model are typically closely related to semantic understanding and textual content. However, our experiments yield the opposite conclusion. High-level semantic information in features provides limited benefits for enhancing transferability.
This phenomenon may relate to the levels of feature abstraction in deep neural networks. At lower levels, the network primarily learns more general, low-level patterns, such as spectral features or time-domain patterns in audio, which have stronger cross-model transferability. On the other hand, higher-level features undergo multiple nonlinear transformations and gradually focus on more complex information, such as semantics and context, which are more specific and context-dependent. Therefore, when adversarial examples transfer from one model to another, lower-level features are more likely to provide stable transfer performance.

\begin{figure}[t]
    \centering
    \includegraphics[width=0.8\linewidth]{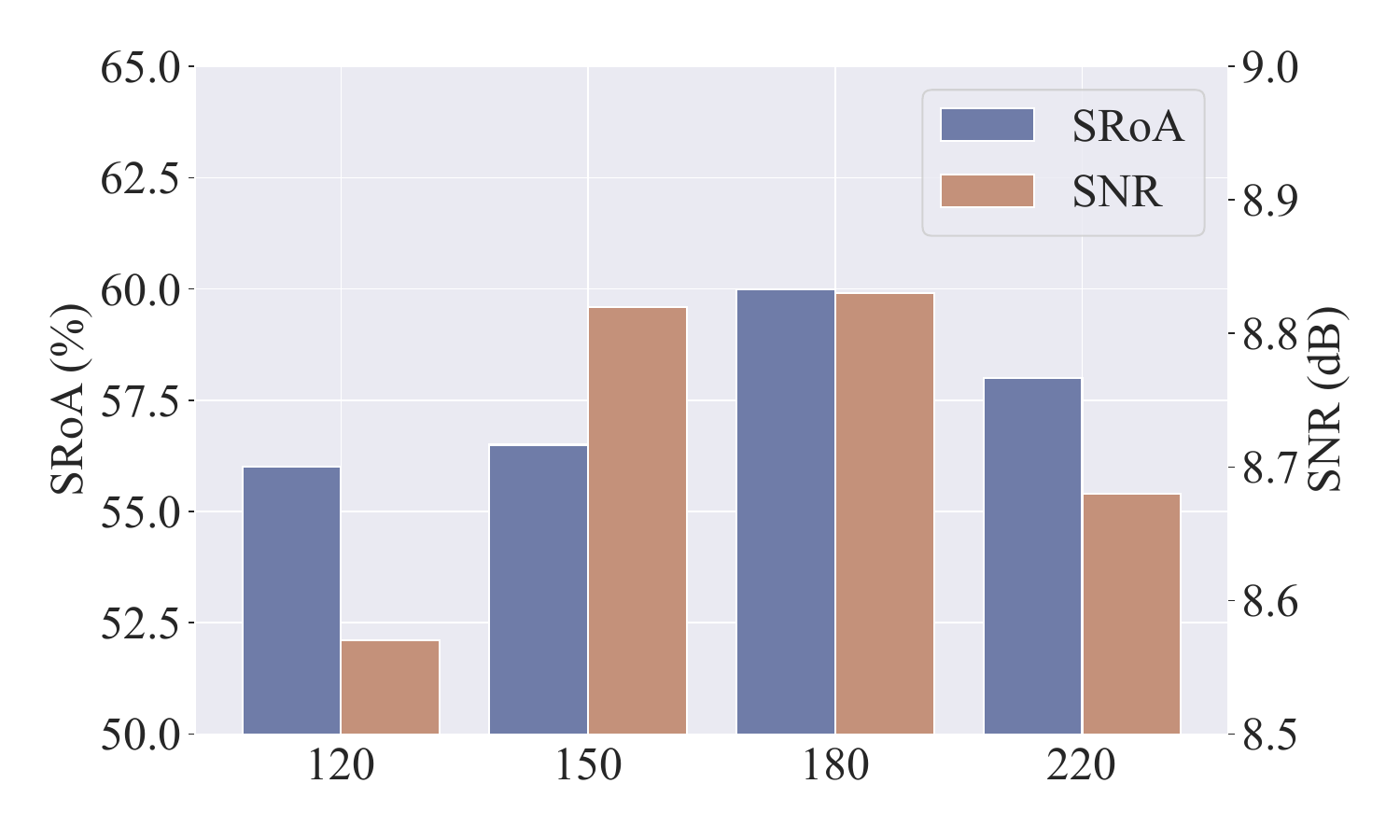}
    \caption{Impact of hyper-parameter $\beta$.}
    \label{fig:factor}
    \vspace{-3mm}
\end{figure}

\subsection{Hyper-parameter Analysis}
A key hyper-parameter in our method is the scaling factor $\beta$, which controls the weight ratio between the two loss functions. Therefore, we designed an ablation experiment to select the appropriate scaling factor. We evaluated four values: 120, 150, 180, and 220, and the experimental results are shown in Figure~\ref{fig:factor}. Overall, our method maintains good attack performance across all scaling factor values, with SRoA consistently above 55\%. Meanwhile, the audio quality does not significantly degrade, and the SNR remains above 8.5. Notably, when the scaling factor is set to 180, both SRoA and SNR reach their highest values. This suggests that, with this value, our method achieves the best attack performance while maintaining excellent audio quality. Therefore, we set the scaling factor to 180.

\section{Conclusion}
In this paper, we propose acoustic representation optimization, which utilizes the lower layers of speech representation models to extract fine-grained acoustic representations. By leveraging the cross-model consistency of lower-level acoustic features, we enhance the transferability of adversarial examples. At the same time, our method is plug-and-play, enabling integration with existing methods to improve their transferability. Experiments on three existing methods across three modern ASRs demonstrate the effectiveness and practicality of our approach, while also ensuring audio quality. Future work will focus on how to improve the transferability of targeted attacks and inspire the development of more robust ASR models.

\section*{Acknowledgment}
This research is supported in part by the Beijing Natural Science Foundation under Grant No. QY24206, and the Fundamental Research Funds for the Central Universities under Grant No. 2024ZCJH05.

\bibliographystyle{IEEEbib}
\bibliography{ref}

\end{document}